\documentclass[a4paper]{jpconf}
\pdfoutput=1
\usepackage{amsmath}
\usepackage{graphicx} 
\usepackage{braket}

\begin{document}

\title{Thermalization in parametrically driven coupled oscillators}

\author{Sayak Biswas and S. Sinha}

\address{Indian Institute of Science Education and
Research-Kolkata, Mohanpur, Nadia-741246, India}

\vspace{10pt}
\begin{indented}
\item[]March 2020
\end{indented}

\begin{abstract}
We consider a system of two coupled oscillators one of which is driven parametrically and investigate both classical and quantum dynamics within Floquet description. Characteristic changes in the time evolution of the quantum fluctuations are observed for dynamically stable and unstable regions. Dynamical instability generated by the parametrically driven oscillator leads to infinite temperature thermalization of the undriven oscillator which is evident from the equi-partitioning of energy, reduced density matrix and saturation of entanglement entropy. We also confirm  that the classical Lyapunov exponent is correctly captured from the growth rate of `unequal time commutator' of dynamical variables which indicates thermalization stems from the underlying dynamical instability in quantum system.   
\end{abstract}

%\maketitle

\section{Introduction}
Formulation of statistical mechanics relies on the important postulate of equivalence between the time average and ensemble average of physical observables\cite{penrose}. A deeper understanding of ergodic hypothesis leads to interesting questions related to thermalization. Dynamical route to thermalization has recently been investigated for various quantum systems\cite{D_Alessio_2016}. Eigenstate thermalization hypothesis has been put forward to understand thermalization of isolated quantum systems\cite{PhysRevA.43.2046,PhysRevE.50.888,reimann} which has close connection with random matrix theory\cite{l.santos}. It is generally believed that the non-integrability and underlying chaos in a system with many degrees of freedom are crucial ingredients for thermalization which results in sharing of energy between many degrees of freedom.
Ergodicity requires the system to explore available phase space volume or microstates with equal probability. For classical system such mixing in phase space can be triggered by underlying chaotic dynamics, whereas such picture is not clear in the case of quantum system due to the absence of phase space trajectories. However in certain systems the connection between thermalization and underlying chaos have been explored\cite{Altland_2012,alt_prl,SRay}. Since dynamical instability can lead to chaos, it is interesting to investigate whether ergodic behavior and thermalization is observered in a dynamically unstable system. 

In this work we study a system of coupled oscillators, one of which is driven parametrically giving rise to dynamical instability for appropriately chosen values of the parameters. Our main aim is to investigate the signature of thermalization in the undriven oscillator from the partitioning of energy between the degrees of freedom and whether the parametrically driven oscillator can act as heat bath for the coupled oscillator in the unstable region.  
 
The Hamiltonian describing the driven coupled oscillator model is given by,
\begin{equation}
\mathcal{H} = \frac{{P_1}^{2}}{2m}+\frac{1}{2}m\omega_{1}^{2}Q_1^{2}+\frac{P_{2}^{2}}{2m}+\frac{1}{2}m\omega_0^2\nu^2(t){Q_2}^{2}+\Lambda{Q_1}{Q_2},
\label{Ham}
\end{equation}
where $Q_{1,2}$,$P_{1,2}$ are position and momentum of first and second oscillator with equal mass $m$. We consider a linear coupling between the positions of the two oscillators with strength $\Lambda$. 
Second oscillator is driven periodically with a time period $T/\omega_0$, so that the time dependent frequency satisfies the condition $\nu(t+T/\omega_0) = \nu(t)$. 
For simplicity we consider a periodic drive in the form of square wave pulse which allows us to perform the calculation analytically. 
The frequency modulation of the second oscillator can be written as,
\begin{equation}
\nu(t)=\begin{cases}
~(1+\epsilon)~~\mbox{for}~ & nT\leq \omega_0t < (n+ \frac{1}{2})T\\
~ (1-\epsilon) ~~\mbox{for}~ & (n+ \frac{1}{2})T \leq \omega_0t < (n+1)T
\end{cases}
\label{sq_drive}
\end{equation}

where $n$ is an integer, $T$ is the time period of the drive in units of $1/\omega_0$ and $\epsilon$ denotes the magnitude of the frequency modulation. For driven oscillator, dynamical instability can be triggered by tuning the parameters $\epsilon$ and $T$. 
In the coupled oscillator model we intend to investigate the dynamical behavior and thermalization of the first oscillator as a result of the dynamical instability induced by the driven (second) oscillator.

The paper is organized as follows; in section 2, we study the dynamics of periodically driven coupled oscillator model using Floquet method. Instability regions of parametrically driven oscillators are identified for different values of driving parameters in subsection 2.1.  
In subsection 2.2, we consider quantum dynamics in terms of time evolution of canonical conjugate operators. From the fluctuation of canonical variables we investigate equipartition of energy as a signature of thermalization process. Identification of instability in quantum dynamics from unequal time commutators is discussed in subsection 2.3. 
Next in section 3, we analyze the reduced density matrix of the undriven oscillator and corresponding entanglement entropy to investigate thermalization for different parameters of driving. Finally we summarize the results in section 4.

\section{Dynamics}
In this section we discuss both classical and quantum dynamics of the driven coupled oscillator model. Classical dynamics of the system is described by the Hamilton's equation of motion,
\begin{equation}
\frac{dQ_i}{dt}=\frac{\partial\mathcal{H}}{\partial P_i};\frac{dP_i}{dt}=-\frac{\partial \mathcal{H}}{\partial Q_i}.
\label{ham_eq}  
\end{equation}
where the index $i=1,2$ describes two oscillators. From now on we use dimensionless variables $q_i = Q_i\sqrt{\frac{m\omega_0}{\hbar}}$, and $p_i = P_i /\sqrt{\hbar m \omega_0}$. Correspondingly time and energy are measured in units of $1/\omega_0$ and $\hbar \omega_0$ respectively. 
Time evolution of these dimensionless phase space variables can be cast into a compact form,
\begin{equation}
\frac{dX_i(t)}{dt} = \sum_j A_{ij} (t)X_j(t),
\label{equation_motion}
\end{equation}
where, $X_1=q_1$, $X_2=q_2$, $X_3=p_1$, $X_4=p_2$, and $A(t)$ is the time dependent matrix,
\begin{equation}
A(t) = \begin{bmatrix}0&0&1&0\\0&0&0&1\\-\alpha^2&-\lambda&0&0\\-\lambda&-\nu^2(t)&0&0\\\end{bmatrix}.
\label{mat_A}
\end{equation}
where $\alpha=\omega_1/\omega_0$,and $\lambda=\Lambda/(m\omega_0^2)$ are the dimensionless parameters. We observe that the main features of the dynamics do not crucially depend on the parameter $\alpha$ and for simplicity, we set $\alpha=1$ in the rest of our analysis.

Time evolution of corresponding quantum operators $\hat{X}_{i}$s can be obtained from Heisenberg equation of motion $\frac{d\hat{X}_{i}(t)}{dt}=-i[H,\hat{X}_{i}(t)]$, where $H$ is the dimensionless Hamiltonian, $\mathcal{H}/\hbar\omega_0$. It is interesting to note that both the quantum operators and corresponding classical dynamical variables satisfy same set of linear equations given in Eq.\ref{equation_motion} since the Hamiltonian is quadratic in $p_{i}$s and $q_{i}$s. 

For periodic drive, the matrix $A(t)$ is also periodic in time which enables us to solve Eq. \ref{equation_motion} using Floquet method\cite{ASENS_1883_2_12__47_0}. The Floquet operator (matrix) $F(T)$ describes the time evolution of both classical as well as quantum dynamical variables $X_{i}(t)$s over a time period $T$\cite{PhysRevA.81.012316}.  
In terms of the Floquet matrix $F$, the stroboscopic dynamics after each interval of time period $T$ can be written as a discrete map of classical as well quantum observables,
\begin{equation}
X_{i}(n+1) = \sum_{j} F_{i,j}(T) X_{j}(n),
\label{evol_map}
\end{equation}
where $X_{i}(n) = X_{i}(t=nT)$ and $n=0,1,2...$. From now on the integer $n$ plays the role of time and we study both the classical and quantum dynamics stroboscopically. 
The choice of driving protocol given in Eq.\ref{sq_drive} simplifies the calculation and allows us to compute the Floquet matrix $F$ analytically. The matrix $A(t)$ is piece wise constant over half of the time period of the square pulse drive and is denoted by $A_{\pm}$ for 
$\nu = 1 \pm \epsilon$. The Floquet matrix is given by,
\begin{equation}
F = e^{TA_{-}/2}e^{TA_{+}/2}.
\label{floquet_eq}
\end{equation}
It is important to note that the Floquet matrix $F$ depends on the parameters of driving $\epsilon$, $T$ as well on the system parameters $\lambda$ and $\alpha$. The dynamical behavior of the system crucially depends on $F$ which can change its nature by tuning the driving parameters $\epsilon$ and $T$. Generally parametrically driven oscillator becomes dynamically unstable for certain range of driving parameters $\epsilon$ and $T$\cite{arnold}. Manifestation of such classical dynamical instability in quantum dynamics is the main focus in the present study. We also investigate whether such instability generated 
by driving the second oscillator can thermalize the first oscillator connected to it.

\subsection{Classical Dynamics}
In this subsection we discuss the classical dynamics of parametrically driven oscillators using Floquet technique mentioned above and identify 
instability regions in parameter space. 

First we review analytically solvable model of single driven oscillator. In absence of the coupling term $\lambda =0$, the second oscillator can be described by the Hamiltonian,
\begin{equation}
H^0(p,q) = \frac{p^2}{2}+ \nu^2(t)\frac{q^2}{2},
\label{ham_osc2}
\end{equation}
where $\nu^2(t)$ is the same square wave pulse given in Eq.\ref{sq_drive}. The corresponding equation of motion is given by,\begin{equation}
\ddot{q}(t)+\nu^2(t)q(t)=0,
\label{Hills_eq}
\end{equation}
which is known as Hill's equation\cite{hill1886}. Following the method described above,denoting $\nu_\pm = 1 \pm \epsilon$, the $2\times 2$ Floquet matrix is given by,
\begin{equation}
F^0=\begin{bmatrix}\cos(\frac{\nu_-T}{2})\cos(\frac{\nu_+T}{2})-\frac{\nu_+}{\nu_-}\sin(\frac{\nu_-T}{2})\sin(\frac{\nu_+T}{2})&\frac{1}{\nu_+}\sin(\frac{\nu_+T}{2})\cos(\frac{\nu_-T}{2})+\frac{1}{\nu_-}\sin(\frac{\nu_-T}{2})\cos(\frac{\nu_+T}{2})\\-\nu_+\sin(\frac{\nu_+T}{2})\cos(\frac{\nu_-T}{2})-\nu_-\sin(\frac{\nu_-T}{2})\cos(\frac{\nu_+T}{2})&\cos(\frac{\nu_+T}{2})\cos(\frac{\nu_-T}{2})-\frac{\nu_-}{\nu_+}\sin(\frac{\nu_+T}{2})\sin(\frac{\nu_-T}{2})\end{bmatrix}
\label{floquet0}
\end{equation}

\begin{figure}[h]
  
  \centering
    \includegraphics[width=0.80\textwidth]{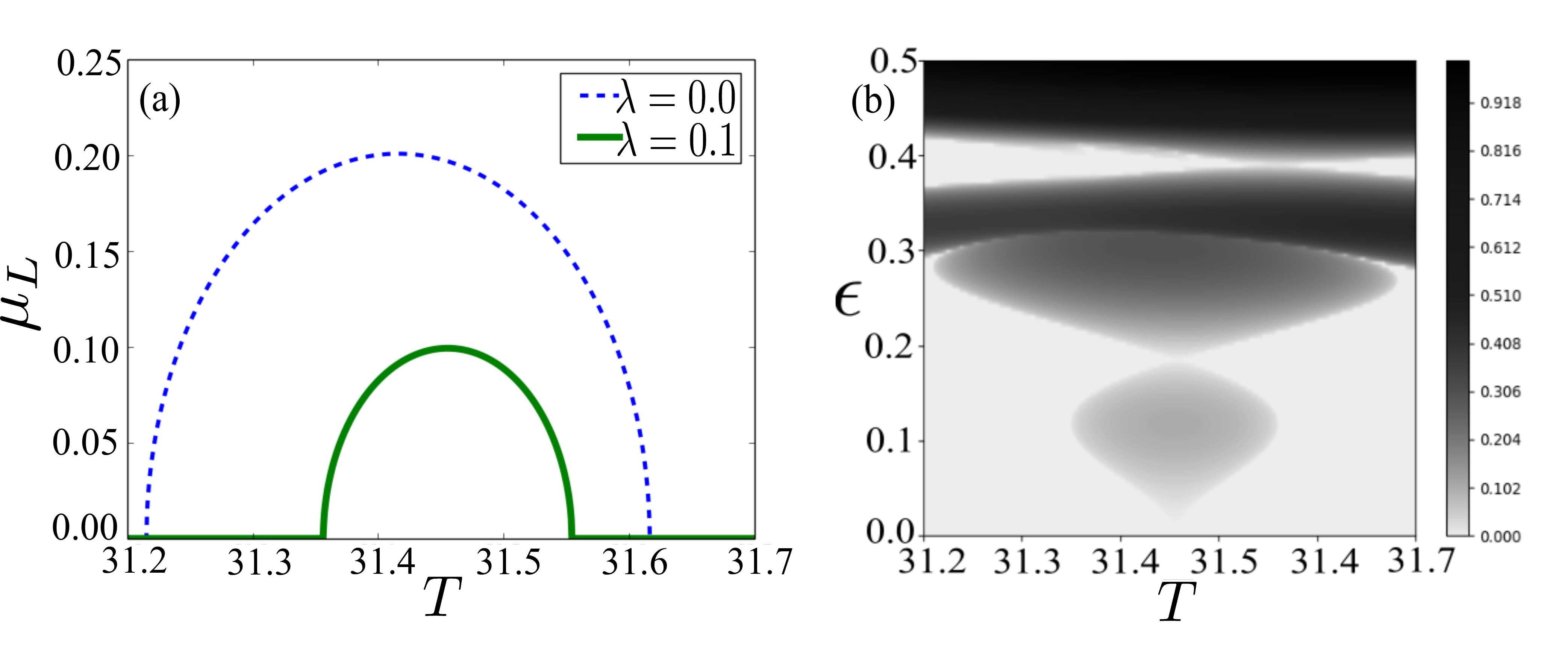}
    \caption{(a)Dependence of $\mu_L$ on $T$ with $\epsilon=0.1$ for decoupled and coupled oscillator.(b)unstable regions with $\mu_L>0$ on  $\epsilon-T$ plane for $\lambda=0.1$.Magnitude of $\mu_L$ is shown by colour-scale. }
    \label{fig1}
\end{figure}

Stroboscopic phase space dynamics can be written in terms of the Floquet matrix,
\begin{numparts}
\begin{eqnarray}
q(n+1) &=& F^0_{11}q(n) + F^0_{12}p(n),\\
p(n+1) &=& F^0_{21}q(n) + F^0_{22}p(n).
\label{strobo0}
\end{eqnarray}
\end{numparts}
Stability of above dynamical map can be determined from the Lyapunov exponent\cite{strogatz} which can be written as,
\begin{equation}
\mu_{L}= \log(|\mu_m|),
\label{lyapunov}
\end{equation}
where $\mu_m$ denotes the eigenvalue of the Floquet matrix $F$ which has largest magnitude. When maximum absolute value of an eigenvalue of the Floquet matrix becomes larger than unity then Lyapunov exponent becomes positive indicating that the fluctuations can grow exponentially with stroboscopic time $n$ leading to dynamical instability. It is important to note that for linear maps like Eq.\ref{floquet0}, Lyapunov exponent and stability can be determined solely from the Floquet matrix. 
In this case $F_{0}$ is $2\times 2$ matrix with eigenvalues $\mu_1$ and $\mu_2$ satisfying the condition $\mu_1 \mu_2 =1$ since the Hamiltonian map preserves phase space area. Magnitude of one of the eigenvalues exceeds unity when $|\mu_1 + \mu_2|\geq 2$. Condition for dynamical instability of parametrically driven single oscillator (described by Eq.\ref{Hills_eq}) is given by,
\begin{equation}
|\cos(T)-\epsilon^2\cos(\epsilon T)|> 1-\epsilon^2.
\label{boundary1}
\end{equation}
Corresponding Lyapunov exponent as a function of time period of drive $T$ for fixed value of $\epsilon$ is shown in Fig.(\ref{fig1}(a)) which vanishes at the boundary given by Eq.\ref{boundary1}.

Next, we repeat this procedure for the original model of coupled oscillator (as in Eq.\ref{Ham}) with $\lambda \neq 0$ and analyze dynamical instability from the largest eigenvalue of the Floquet matrix $F$. The Lyapunov exponent in this case is depicted in Fig.\ref{fig1}(a). In comparison with the single oscillator we observe that the magnitude of Lyapunov exponent and instability region of coupled oscillator changes due to its dependence on the couplings $\lambda$. Also the instability region in $\epsilon - T$ plane obtained from the condition $\mu_{L} >0$ is shown in Fig.\ref{fig1}(b)

\begin{figure}[h!]
  
  \centering
    \includegraphics[width=0.70\textwidth]{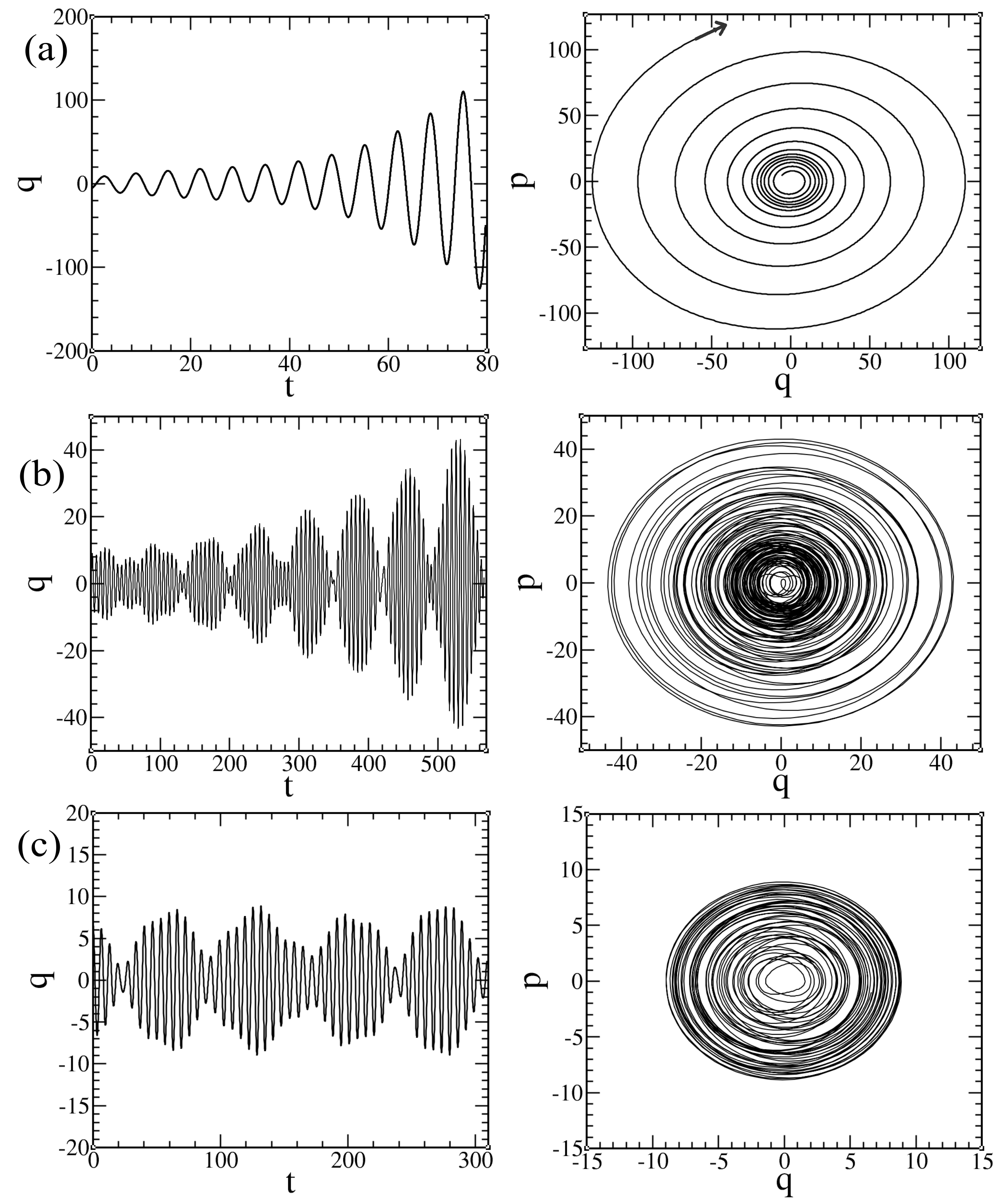}
    \caption{Time evolution of dynamical variable (left) and phase space trajectories (right) corresponding to the first oscillator for :(a)$\mu_L>0$ with real maximum eigenvalue; (b)$\mu_L>0$ with complex maximum eigenvalue; (c) $\mu_L=0$ }
    \label{fig2}
\end{figure}

We can classify the dynamics in three different regions based on the nature of the largest eigenvalue $\mu_m$ of Floquet matrix.
When the magnitude of largest eigenvalue is unity the dynamics is stable since $\mu_{L}=0$. As seen from Fig.\ref{fig2}(c) the first oscillator exhibits oscillatory motion in stable region with beating effects due to the presence of more than one frequency. In the unstable regime with $\mu_{L} >0$ the phase space trajectories spiral out and amplitude of oscillation increases with stroboscopic time, however beating phenomenon persists for complex  $\mu_m$ and complicated phase space trajectory is observed as seen in Fig.\ref{fig2}(b). On the other hand, for real maximum eigenvalue with magnitude larger than unity($\mu_L>0$), the amplitude increases monotonically and phase space trajectory represents as shown in Fig.\ref{fig2}(a). Our goal is to study the manifestation of such instability in quantum dynamics which is discussed in next section.

\subsection{Quantum Dynamics}
In what follows, we investigate the full quantum dynamics of the model Hamiltonian Eq.\ref{Ham} for different frequencies and amplitude of driving. As discussed earlier, Heisenberg evolution of operators, $\hat{X}(t) =\{\hat{q}_i(t),\hat{p}_{i}(t)\}$ can be obtained stroboscopically, in terms of the operators at the initial time $\hat{X}(0)$ using classical Floquet operator $F$.

The expectation values of relevant operators can be obtained from the initial state $|\psi_{i}\rangle$. For simplicity we choose $|\psi_{i}\rangle = \ket{0,0}$ which is the ground state of two decoupled  oscillators in absence of driving. We study the fluctuations of the dynamical variables $\hat{p}_{1}, \hat{q}_{1}$ of the first oscillator to understand degree of chaos generated by its coupling with the driven oscillator. Although the observables are local to the first subsystem, the coupled dynamics entangles the subsystems to produce non trivial behaviour for the quantity, 
\begin{equation}
R(t)= \frac{\langle p_1^2(t)\rangle}{\langle q_1^2(t)\rangle}.
\label{ratio_R}
\end{equation}
Numerically it is observed that the long time behaviour of $R(t)$ is highly sensitive to the parameters we choose for our Hamiltonian. In fact the long time dynamics of this quantity shows distinct characteristics, depending on whether the system is in the regular or unstable region of the parameter space of corresponding classical dynamics. Moreover the long time behavior of $R(t)$ carries the signature of thermalization which will be discussed later. 
It is worthwhile to investigate the evolution of $R$ for long time, which can be obtained analytically as follows.
Using Eq.\ref{evol_map} we obtain,
\begin{equation}
     X_i(n)=\sum_j (F^n)_{ij} X_j(0),
\label{X_n}     
\end{equation}
which yields,
\begin{equation}
     \langle X_i^2(n)\rangle =\bra{0,0}X_i^2(n)\ket{0,0}=\frac{1}{2} \sum_j (F^n)_{ij}^2 ,
\label{sqX_n}          
\end{equation}
Where we used the fact that $\bra{0,0}p_iq_j+q_jp_i\ket{0,0}=0$ and that $\bra{0,0}p_i(0)^2\ket{0,0}=\bra{0,0}q_i(0)^2\ket{0,0}=1/2$. As result the ratio $R(n)$ for large stroboscopic time $n$ can be written as,
\begin{equation}
R(n)=\frac{[F^n(F^T)^n]_{33}}{[F^n(F^T)^n]_{11}},
\label{R_N}
\end{equation}
Where $F^T$ denotes the transpose of F.
 
\begin{figure}[h!]
  
  \centering
    \includegraphics[width=0.7\textwidth]{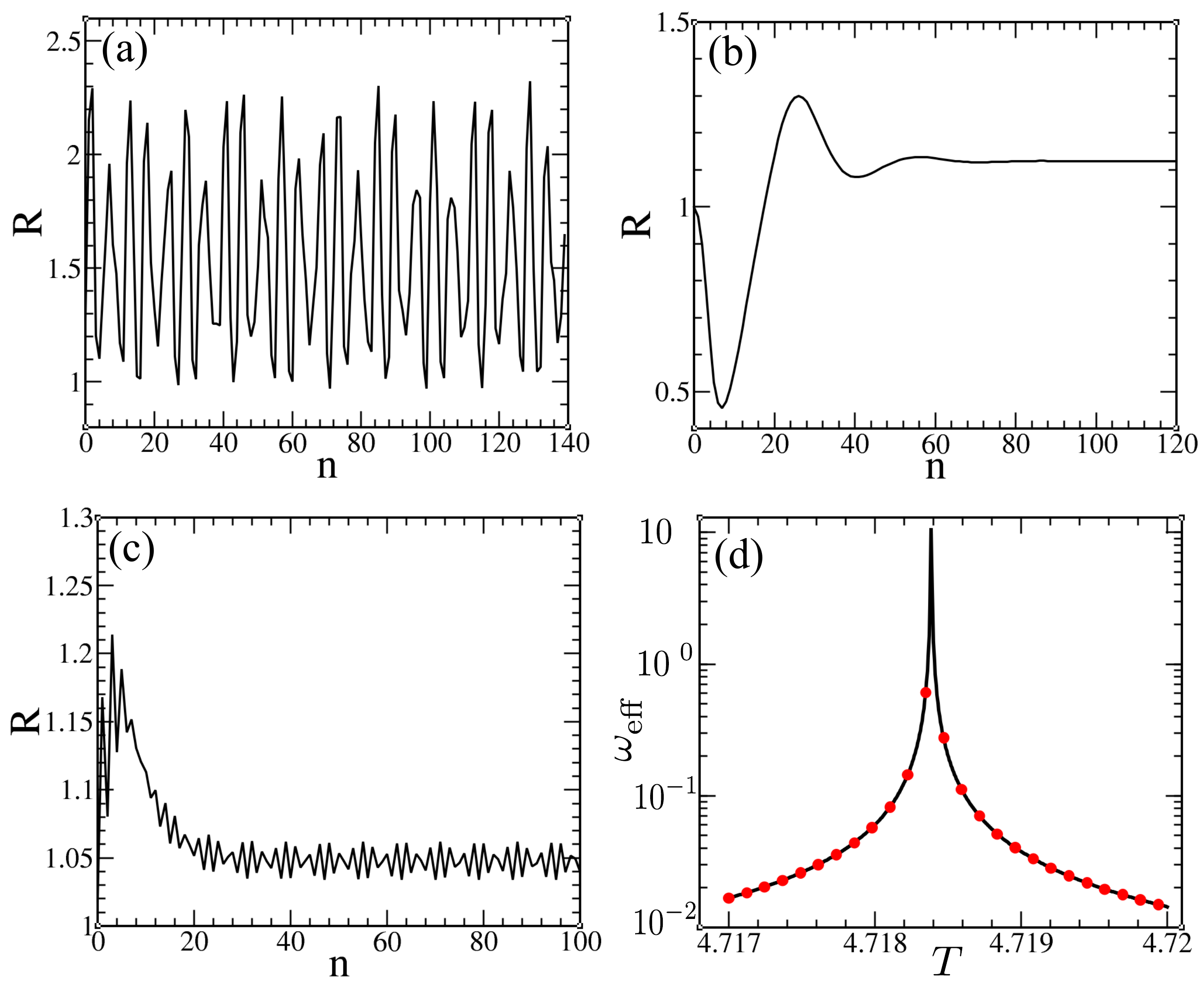}
    \caption{Stroboscopic time evolution of $R$ in three different cases :(a)stable region with  $\mu_L=0$ ,(b)unstable region with $\mu_L>0$ and real $\mu_m$ (c) $\mu_L>0$ with $\mu_m$ complex.(d)In the unstable region, dependence of $\omega_{eff}$  on driving period T for $\lambda=0.1$ and $\epsilon=0.1$. Analytical expression of $\omega_{eff}$ from Eq.(\ref{om_ef}) is denoted by solid line and numerical values obtained from long time dynamics denoted by(red) dots. }
    \label{fig3}
\end{figure}
Next, we study the stroboscopic evolution of the ratio $R(n)$ for sufficiently large $n$ which exhibits three different scenarios by changing the parameters of the drive. For dynamically stable region $R$ exhibits oscillatory behavior as shown in Fig.\ref{fig3}(a). On the other hand it saturates to a steady value in the dynamically unstable regime with 
Lyapunov exponent $\mu_{L} > 0$ which is evident from Fig.\ref{fig3}(b) and (c). However a small oscillation around the steady value of $R$ persists when the largest magnitude eigenvalue of the Floquet matrix $F$ is complex (see Fig.\ref{fig3}(c)).
The saturation of ratio between $\langle \hat{p}_{1}^2 \rangle$ and $\langle \hat{q}_{1}^2 \rangle$ clearly indicates equipartition of energy in the undriven oscillator as a result of dynamical instability generated by the parametrically driven oscillator. In this regime we can consider the undriven oscillator as an isolated system at thermal equilibrium with an effective frequency $\omega_{eff}$ so that 
\begin{equation}
\bar{R} = \langle \hat{p}_{1}^2 \rangle / \langle \hat{q}_{1}^2 \rangle = \omega_{eff}^2
\end{equation}
where $\bar{R}$ is the saturation value of $R(n)$ after sufficiently long time. Using Eq.\ref{R_N} the effective frequency $\omega_{eff}$ can be written in terms of the eigenvector $u$ with largest eigenvalue of  $F^T$, 
\begin{equation}
\omega_{eff}^2 = \frac{|u_3|^2}{|u_1|^2}
\label{om_ef}
\end{equation}
where, $u_1$ and $u_3$ are the elements of the vector $u$ corresponding to the variables $q_1$ and $p_1$ respectively.

This phenomena related to equipartition of energy in undriven first oscillator is an evidence of thermalization which is a consequence of the dynamical instability generated by the driven oscillator. However, thermalization in usual sense is described by the emergence of a steady state corresponding to appropriate statistical ensemble. Unlike thermalization of an oscillator in the presence of a heat bath\cite{Weiss,Hanggi}, in the present case the fluctuations in $p$ and $q$ increase monotonically due to constant pumping of energy in absence of dissipation, while saturation of the ratio between them ($R$) leads to equipartitioning of energy.

\subsection{Unequal Time Commutator }
In recent years, out of time ordered correlators have become an important marker to diagnose underlying instability and chaos in quantum mechanical systems\cite{Maldacena2016}, which is inherently difficult due to the absence of classical trajectories. 
Classically, a dynamical instability, can be quantified by the sensitivity of the trajectory to initial condition, $\frac{\delta q(t)}{\delta q(0)}$ which grows as $e^{\mu_L t}$ at large times, where $\mu_L$ is the Lyapunov exponent. By introducing the poisson bracket, this can be rephrased as
\begin{equation}
    \frac{\delta q(t)}{\delta q(0)}=[q(t),p(0)]_{PB} \sim e^{\mu_L t}
    \label{classical_OTOC}
\end{equation}
For a quantum system one can replace the poisson bracket by the commutator and define the unequal time commutator of an observable $O$,
\begin{equation}
    C(t,t')=-\bra{\psi}[O(t),O(t')]^\dagger[O(t),O(t')]\ket{\psi}
\label{otoc_gen}
\end{equation}
as a quantity for detecting signature of chaos in the context of quantum mechanics. In the semi-classical limit, at small times, one observes an exponential time dependence of this quantity, and the growth rate is a measure for the underlying chaos. But for larger times, there could be significant deviation from this behaviour\cite{Hashimoto2017}.\ 

\begin{figure}[h!]
\centering
\includegraphics[width=0.60\textwidth]{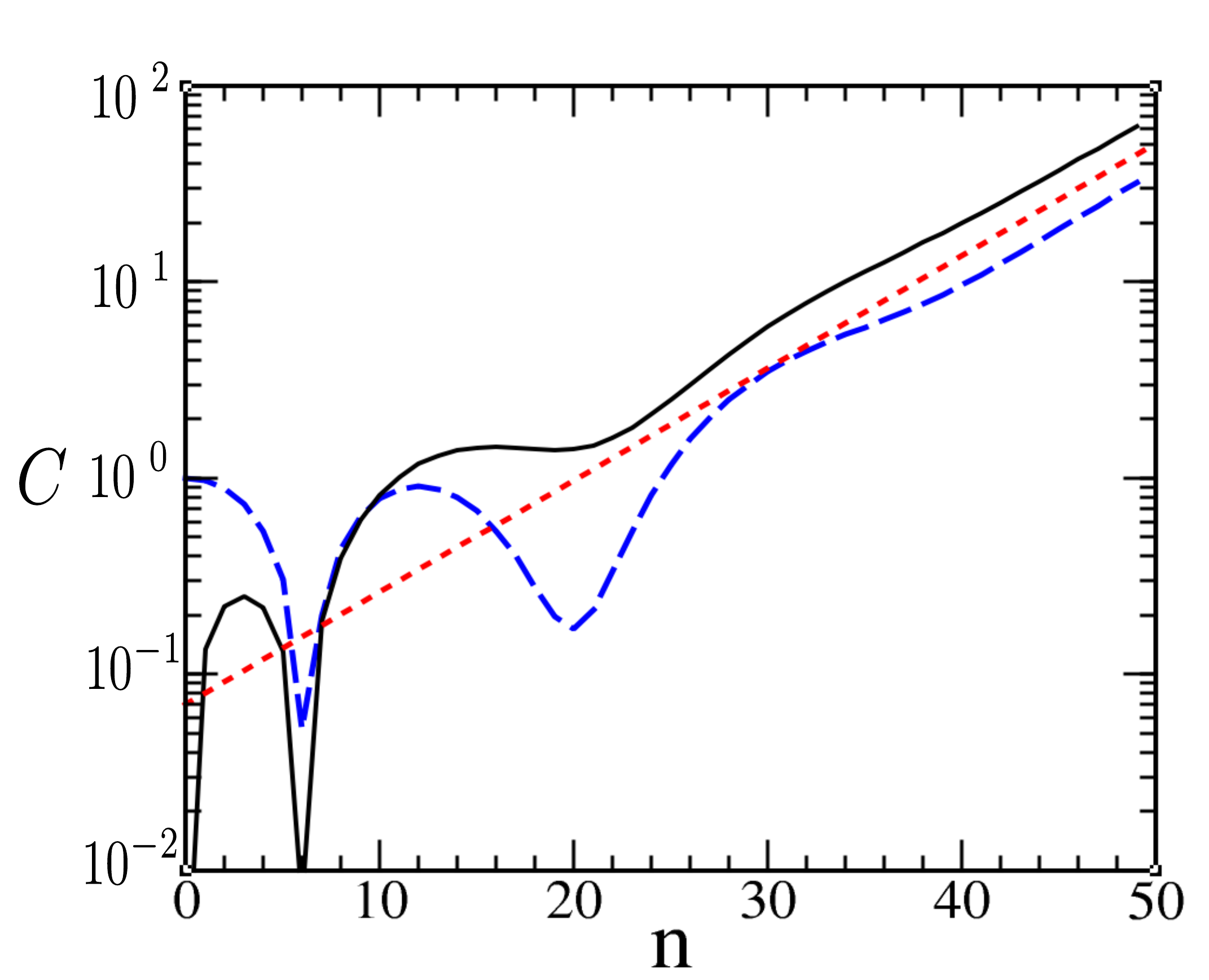}
\caption{Growth of unequal time commutators $C_{qq}$(solid black line) and $C_{pq}$(blue dashed line) for $\lambda=0.1$, $\epsilon=0.1$, $T=3.32$ where $\mu_L>0$. The straight line in (red) dots represents the common slope which corresponds to 2$\mu_L$ }
\label{fig4}
\end{figure}

For the system under study, we study the dynamics of the two quantities given below, stroboscopically using Floquet dynamics
\begin{numparts}
\begin{eqnarray}
C_{qp}(t)&=&-\bra{\psi_i}[p_1(t),q_1(0)]^\dagger[p_1(t),q_1(0)]\ket{\psi_i}  \\
C_{qq}(t)&=&-\bra{\psi_i}[q_1(t),q_1(0)]^\dagger[q_1(t),q_1(0)]\ket{\psi_i} 
\label{OTOC}
\end{eqnarray}
\end{numparts}
A semi-logarithmic plot, as in Fig(\ref{fig4}), shows that in the $\mu_L>0$ region, after some initial fluctuation, the long time behaviour of both these quantities is linear. The common slope captures the classical Lyapunov exponent $\mu_L$. Because our Hamiltonian is quadratic and quantum and classical dynamical quantities follow the same evolution equations, we retrieve the classical behaviour mentioned earlier, exactly, in long time.

\section{The Density Matrix}
As mentioned earlier, in dynamics we choose the initial state $|0,0\rangle$ as the product state of corresponding oscillators ground state which is disentangled system. During time evolution we investigate  how the entanglement between the oscillators grows from the reduced density matrix of the undriven oscillator and corresponding entanglement entropy. 
%So initially we have an disentangled system. With time, the two oscillators get entangled owing to the coupling and the drive. Our aim in this section is to study how this entanglement behaves across different dynamical regions identified earlier. 
This will eventually lead us to a deeper understanding of the onset of thermalization in this system.

For above mentioned initial state the dynamics of the full system can be described by the following ansatz of the wave function,
\begin{equation}
\Psi(q_1,q_2,t)= N(t)exp(-\frac{a(t)q_1^2+b(t)q_2^2+2c(t)q_1q_2}{2})
\label{wavefunc}
\end{equation}
where $a(t),b(t),c(t),N(t)$, are complex time dependent parameters of the normalized wave function. From the time dependent Schr\"odinger equation we obtain following set of equations describing time evolution of the parameters,
\begin{numparts}
\begin{eqnarray}
&&i\dot a=a^2+c^2-1  \\
&&i\dot b=b^2+c^2-\nu^2(t) \\
&&i\dot c=ac+bc-\lambda \\
&&i\dot N=N(a+b)
\label{param_eq}
\end{eqnarray}
\end{numparts}
with the initial condition fixed by the initial state.
The density matrix $\hat{\rho}$ of the total system can be constructed from the above mentioned time dependent wave function which is essentially a pure state and in the coordinate representation it is given by,
\begin{equation}
\rho(q_1',q_2',q_1,q_2,t)=\Psi^{*}(q_1',q_2',t)\Psi(q_1,q_2,t).
\label{rho_tot}
\end{equation}
However our interest lies in the reduced density matrix of the first oscillator, which can be obtained by tracing out the degrees of freedom of the second oscillator. Again in coordinate representation the reduced density matrix $\hat{\rho}_1$ corresponding to the undriven (first) oscillator can be written as,
\begin{equation}
\rho_1(q_1',q_1,t)=\int{dq_2   \rho(q_1',q_2,q_2,q_1,t)}
\label{rho_red}
\end{equation}
From the wavefunction given in Eq.\ref{wavefunc} the reduced density matrix at time $t$ is given by,
\begin{equation}
\rho_1(q_1',q_1,t)=|N(t)|^2\sqrt{\frac{2\pi}{\mu(t)}}{exp(-\frac{\chi(t) q_1^2+\chi(t)^*q_1'^2-2\eta(t) q_1'q_1}{2})}
\label{rho_red_form}
\end{equation}
Where,
\begin{equation}
\mu(t)=2Re(b(t)),~~
\eta(t)=\frac{|c(t)|^2}{\mu(t)},~~
\chi(t)=a(t)-\frac{c^2(t)}{\mu(t)},~~
\label{rho_red_param}
\end{equation}
describe the time evolution of $\hat{\rho}_1$. 
Since initially we start with a product state, the reduced density matrix of the first oscillator is one that of a pure state with $Tr(\hat{\rho_1}^2)=1$ at the initial time. However during time evolution 
both the oscillators become entangled and the reduced density matrix represents a mixed state with $Tr(\hat{\rho_1}^2)<1$. In other words, the bipartite entanglement entropy becomes non vanishing as the system evolves.

\begin{figure}[h!]
  \centering
    \includegraphics[width=0.85\textwidth]{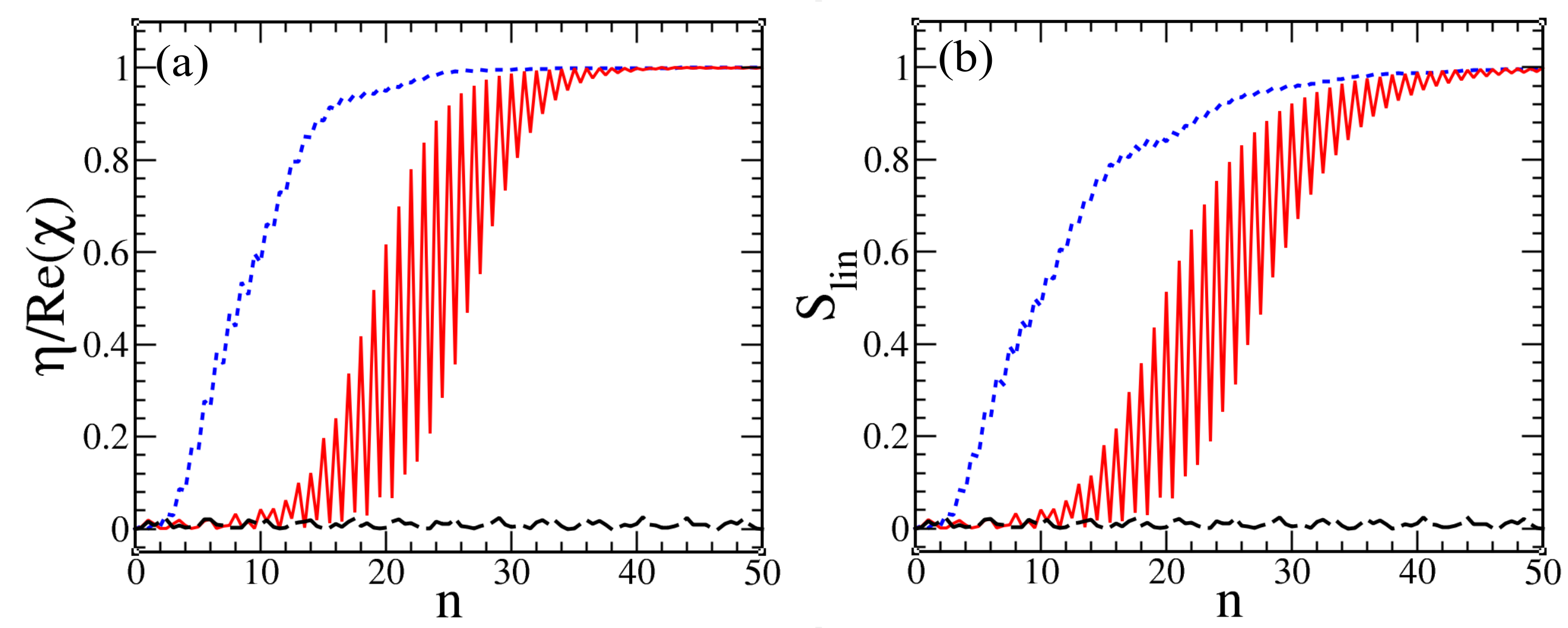}
    \caption{Stroboscopic evolution of (a) density matrix parameter $\frac{\eta}{Re(\chi)}$ indicating thermalization, (b) linear entropy, $S_{lin}$ for three dynamical regimes : $\mu_L=0$ ( black dashes), $\mu_L >0$ and $\mu_m$ complex (solid red), $\mu_L >0$ and $\mu_m$ real(blue dots) }
    \label{fig5}
\end{figure}

Moreover the long time behaviour of the reduced density matrix of the subsystem is of interest to us since we expect thermalization in the dynamically unstable region as indicated by the equipartitioning of energy. We ask the question whether in the long time limit $\rho_1$ resembles the equilibrium density matrix of an isolated oscillator as a result of thermalization. 
The equilibrium density matrix of an oscillator with angular frequency $\omega_{eff}$ at inverse temperature $\beta$ is given by\cite{feynman},
\begin{equation}
\rho_{\beta} = \frac{\sum_{n=0}^{\infty}e^{-n\beta\omega_{eff}}{\braket{q_1'|n}\braket{n|q_1}}}{Z(\beta)}=\mathcal{N(\beta)}\exp[ cosech(\beta\omega_{eff}) q_1'q_1 - \coth(\beta\omega_{eff})(q_1^2+q_1'^2)/2]
\label{rho_therm}
\end{equation}
As a consequence of thermalization we expect,
\begin{equation}
\lim_{t \to \infty}\rho_{1}(q_1',q_1,t) = \rho_{\beta}.
\label{therm_dm}
\end{equation}
Comparing Eq.\ref{rho_red_form} with Eq.\ref{rho_therm} , the inverse temperature $\beta$ can be obtained from the steady value of the 
parameters $\chi(t)$ and $\eta(t)$ in the long time limit.
From numerical calculations we observe  
in the stable regime where $\mu_L=0$, the system does not attain a steady state after long time and all parameters of the reduced density matrix exhibit oscillatory behavior with time. 
On the contrary, $\rho_{1}(q_1',q_1,t)$ attains a steady state in the dynamically unstable regime characterised by positive Lyapunov exponent, however at large time, both the parameters $\chi$ and $\eta$ increase whereas $\eta/Re(\chi)$ saturates to unity, as shown in fig.(\ref{fig5}(a)).
Comparing Eq. \ref{rho_red_form} and \ref{rho_therm}, we can immediately conclude that the subsystem equilibrates effectively at infinite temperature corresponding to $\beta=0$. 
Thermalization to infinite temperature corresponds to microcanonical ensemble represented by diagonal ensemble with equal probability. 
Another quantity of interest is the bipartite entanglement entropy\cite{nielsen}. Here we calculate the linear entropy using the reduced density matrix of the first subsystem,
\begin{equation}
S_{lin}(t)= 1- Tr(\rho_1^2(t)).
\label{lin_entropy}
\end{equation}
which can be calculated stroboscopically from the density matrix parameters as,
\begin{equation}
    S_{lin}(t)=1-\frac{|N(t)|^4~ 2\pi^{2}}{\mu(t)\sqrt{Re(\chi(t))^2-\eta^2(t)}}
    \label{lin_ent_form}
\end{equation}
This quantity, once again, saturates to unity in the dynamically unstable regions (shown in Fig.\ref{fig5}(b)) as the second term in Eq.\ref{lin_ent_form} vanishes with time.

For a system in microcanonical ensemble the linear entropy takes the value $1-\frac{1}{d}$, where d is the dimension of the Hilbert space. 
Once again the saturation of $S_{lin}$ to unity confirms infinite temperature thermalization corresponding to microcanonical ensemble.\cite{rigol_entangle}
Such behavior has also been observed for other quantum systems in the presence of periodic driving\cite{drive,SRay,kapitza}. This simple model provides a clear picture of thermalization and its connection with dynamical instability triggered by driving the system periodically. 
Generally, a subsystem thermalizes in the presence of a heat bath consisting of many degrees of freedom, however in the present model single driven oscillator can act as a reservoir for undriven oscillator, giving rise to thermalization in the dynamical unstable region.\\

\begin{figure}[h!]
  \centering
    \includegraphics[width=0.85\textwidth]{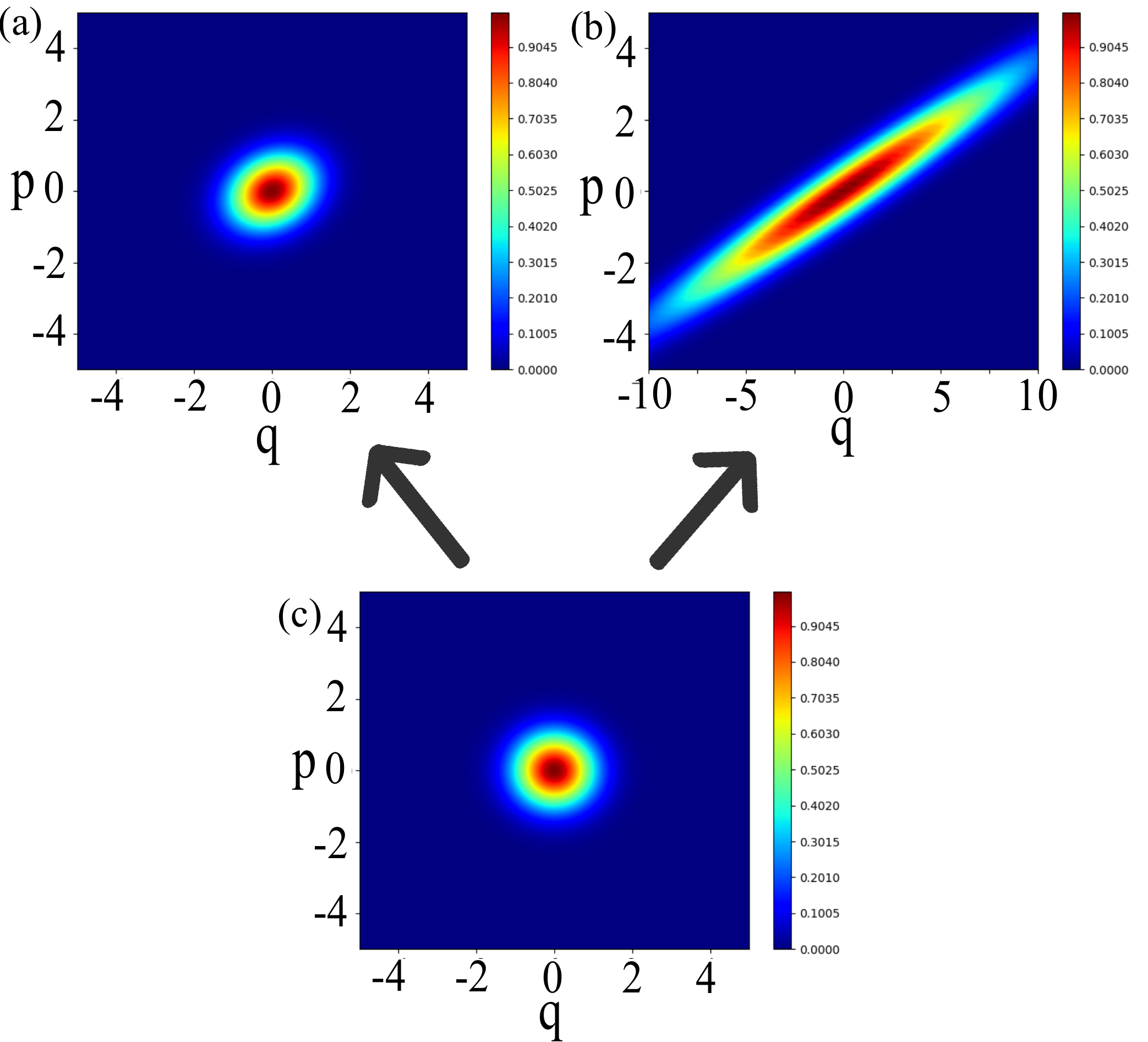}
    \caption{Long time behaviour of Wigner function (a) dynamically stable region,(b) dynamically unstable region, (c) initial  Wigner function }
    \label{fig6}
\end{figure}

The dynamics can be understood better by studying the time evolution of the Wigner function corresponding to $\rho_1$ \cite{wigner}. The Wigner function, 
\begin{equation}
    W(p,q,t)= \frac{1}{2\pi}\int e^{ips}\rho_1(q-s/2,q+s/2,t) \,ds
\end{equation}{}

can be interpreted as a quasi-probability distribution over the phase space of the first oscillator. This is possible because  $\rho_1$ is Gaussian, which ensures positivity of $W$ \cite{wigner_pos}. Initially, the Wigner function is fairly localized (see fig.\ref{fig6}(c)) in the phase space. When the system is being driven in the dynamically unstable region, as time progresses, $W$ spreads over the phase space (as shown in fig.\ref{fig6}(b)). The spreading is however different along the two axes. The long time ratio of the variances along the axes, $\langle p^2 \rangle/\langle q^2 \rangle$ approaches the same $\omega^2_{eff}$ that was discussed in section 2.2. In the dynamically stable region, the Wigner function however remains characteristically localized even at large time and only shows mild oscillations(fig.\ref{fig6}(a)). Present model has been generalized to describe parametrically driven oscillator coupled to a heat bath consisting of many harmonic oscillators, which also exhibits increasing amplitude of fluctuations in the dynamically unstable region, whereas in the stable region they oscillate around the equilibrium value\cite{Hanggi_parametric}. 
    
\section{Conclusions}
To summarize, we have studied dynamics of a coupled driven oscillator and elucidate how dynamical instability of the driven oscillator leads to thermalization of the undriven oscillator. In this simple model both the classical and quantum dynamics has been studied using classical Floquet description. The instability regions in the space of driving parameters are identified by positive Lyapunov exponent calculated from the Floquet matrix. In quantum dynamics, we observe contrasting behavior of fluctuations of canonical coordinates in stable and unstable regions. In the dynamically unstable region the fluctuations of position and momentum of the undriven oscillator grow linearly with stroboscopic time giving rise to equipartitioning of energy. It has also been demonstrated that  due to the quadratic nature of the Hamiltonian, the  exponential growth of the unequal time commutators of the dynamical variables corresponding to the undriven oscillator correctly captures the Lyapunov exponent. We also investigated the growth of entanglement from the reduced density matrix and the corresponding Wigner function of the undriven oscillator. In the dynamically unstable region linear entropy saturates to unity indicating thermalization to infinite temperature which also becomes evident from the reduced density matrix of undriven oscillator.

In conclusion, this simple model elucidates how a parametrically driven oscillator with only one degree of freedom can act as a thermal reservoir for another oscillator connected to it, which thermalizes at infinite temperature corresponding to microcanonical ensemble.

\subsection*{Acknowledgements}
We thankfully acknowledge Mr. Sudip Sinha for discussion.

{}

\end{document}